\documentclass{aastex631}
\usepackage{amssymb}
\usepackage{bm}
\usepackage{times}
\usepackage{amsmath}
\usepackage{gensymb}
\usepackage{graphicx}
\usepackage{xcolor}

\newcommand{\beq}{\begin{equation}}
\newcommand{\kms}{km~s$^{-1}$}
\newcommand{\eeq}{\end{equation}}

\newcommand{\cm}{cm$^{-2}$}

\newcommand{\Msun}{\textrm{M}_\odot}
\newcommand{\kmps}{km~s$^{-1}$}

\newcommand{\hi}{H{\sc i}}
\newcommand{\hii}{H{\sc i} 21\,cm}

\shorttitle{An FRB progenitor born in a galaxy merger}
\shortauthors{Kaur et al.}
\begin{document}
\title{A fast radio burst progenitor born in a galaxy merger}
\correspondingauthor{Nissim Kanekar}
\email{nkanekar@ncra.tifr.res.in}

\author{Balpreet Kaur}
\affiliation{National Centre for Radio Astrophysics, Tata Institute of Fundamental Research, \\Pune 411007, India}

\author{Nissim Kanekar} 
\affiliation{National Centre for Radio Astrophysics, Tata Institute of Fundamental Research, \\Pune 411007, India}

\author{J. Xavier Prochaska}
\affiliation{University of California, 1156 High Street, Santa Cruz, CA 95064, USA}
\affiliation{Kavli Institute for the Physics and Mathematics of the Universe,
The University of Tokyo, 5-1-5 Kashiwanoha, Kashiwa, 277-8583, Japan}

\begin{abstract}
	We report a Giant Metrewave Radio Telescope H{\sc i} 21\,cm mapping study of the neutral atomic hydrogen (H{\sc i}) in the host galaxy of the fast radio burst (FRB) FRB\,20180916B at $z \approx 0.03399$. We find that the FRB host has an H{\sc i} mass of $\rm M_{H{\textsc i}} = (2.74 \pm 0.33) \times 10^9 \, M_\odot$ and a high H{\sc i}-to-stellar mass ratio, $\approx 1.3$. The FRB host is thus a gas-rich but near-quiescent galaxy, that is likely to have acquired  a significant mass of H{\sc i} in the recent past. The H{\sc i} distribution  is disturbed, with extended H{\sc i} 21\,cm emission detected in a north-eastern tail, a counter-tail towards the south, an H{\sc i} hole between the galaxy centre and the FRB location, and a high H{\sc i} column density measured close to the FRB position. The FRB host is part of a group with four companions detected in their H{\sc i} 21\,cm emission, the nearest of which is only 22~kpc from the FRB location. The gas-richness and disturbed H{\sc i} distribution indicate that the FRB host has recently undergone a minor merger, which increased its H{\sc i} mass, disturbed the H{\sc i} in the galaxy disk, and compressed the H{\sc i} near the FRB location to increase its surface density. We propose that this merger caused the burst of star-formation in the outskirts of the galaxy that gave rise to the FRB progenitor. The evidence for a minor merger is consistent with scenarios in which the FRB progenitor is a massive star, formed due to the merger event.
\end{abstract}


\section{Introduction}

The origin of fast radio bursts \citep[FRBs; ][]{Lorimer2007}, the cosmologically-distant, highly energetic, milli-second radio pulses, is a mystery today \citep[e.g.][]{Cordes2019}. 
Understanding FRBs requires understanding the environments in which the bursts form; this is possible for FRBs that have been localized to $\lesssim 1$~kpc accuracy \citep[e.g.][]{Marcote2017,Marcote2020,Chittidi2020,Kirsten2021,Nimmo2021}, allowing one to study the FRB environment in detail. Indeed, the nature of the FRB environments and of the host galaxies of FRBs are among the critical outstanding questions in the field.

Studies of FRB host galaxies have so far focussed on the stellar component of the FRB hosts \citep[e.g.][]{Heintz2020,Tendulkar2021,Mannings2021}. We know very little about the gas conditions that triggered the star formation that is likely to have resulted in the FRB activity. Mapping the gas kinematics is critical for such understanding, as processes that give rise to violent star formation often leave indelible imprints upon the neutral gas. \hii\ mapping studies provide an especially powerful tool to understand the recent history of a galaxy, via the signatures of different physical processes (e.g. smooth rotation, interactions, mergers, outflows, etc) in the gas spatial or velocity distributions \citep[e.g.][]{Sancisi2008}. This is especially important in the case of minor mergers (i.e. between a massive galaxy and a smaller satellite), as these are difficult to detect in the optical photometry of galaxies, but may be identified as disturbances in the gas distribution. For example, \hii\ mapping studies of the gas kinematics in the host galaxies of the nearest gamma ray burst, GRB~980425, and the fast luminous transient AT2018cow have yielded clear evidence for a disturbed gas disk, due to a merger, suggesting that the merger triggered massive star formation activity and hence the formation of the transient \citep[e.g.][]{Arabsalmani2015,Arabsalmani2019,Roychowdhury2019}.
Unfortunately, at present, very few identified FRB host galaxies are at sufficiently low redshifts, $z \lesssim 0.1$, to allow such sensitive \hii\ mapping studies.

FRB\,20180916B is a source of repeating bursts, recently discovered by the CHIME telescope \citep{Chime2019}. Radio interferometric observations of FRB\,20180916B have shown that the source is located in the outskirts of a spiral galaxy, SDSS J015800.28+654253.0 (hereafter, SDSS~J0158+6542), at $z = 0.03399$ \citep{Marcote2020}. While no optical continuum or H$\alpha$ emission is detected at the FRB location, the FRB lies $\approx 250$~pc from the brightest pixel of a region of active star-formation \citep{Marcote2020,Tendulkar2021}. The H$\alpha$ emission from the FRB host is well fit by a thin rotating disk model (with an inclination of $33^\circ$), as expected for a spiral galaxy \citep{Tendulkar2021}. 

We have used the Giant Metrewave Radio Telescope (GMRT) to carry out a deep \hii\ study of SDSS~J0158+6542, aiming to map its \hii\ emission and determine its \hi\ properties. In this {\it Letter}, we describe the GMRT results, that have yielded the first image of the \hi\ distribution in an FRB host galaxy.\footnote{We use a standard flat, $\Lambda$-cold dark matter cosmology throughout this paper, with $\Omega_\Lambda = 0.685$, $\Omega_m = 0.315$, and H$_0 = 67.4$~km~s$^{-1}$~Mpc$^{-1}$ \citep{Planck2020}.}

\section{Observations and data analysis}
\label{sec:obs}

\begin{table}
\centering
\begin{tabular}{|c|c|c|c|c|}
\hline
\hline
UV$_{\rm max}$	& Resolution 	&   P.A. &  RMS noise        &  N$_{\rm {H\textsc{i}}}$~($3\sigma$) \\ 
(k$\lambda$) & ($'' \times ''$) & $^\circ$       & (mJy beam$^{-1}$) &  ($\times 10^{20}$~cm$^{-2}$) \\
\hline
$3.5$ 	& $49.7 \times 38.5$  	&  $7$   &  $0.72$      &  $0.16 $ \\
\hline
$8.0$ 	& $32.0 \times 32.0$  	&  $0$   &  $0.59$      &  $0.24$  \\
\hline
$15$ 	& $15.5 \times 15.5$  	&  $0$   &  $0.40$      &  $0.70$ \\
\hline
$25$ 	& $9.0 \times 9.0$  	&  $0$   &  $0.32$      &  $1.6$  \\
\hline
$40$  	& $5.6 \times 4.7$    	&  $29$  &  $0.27$      &  $4.2$  \\
\hline
$60$  	& $3.8 \times 3.2$    	&  $16$  &  $0.23$      &  $7.7$  \\
\hline
\hline
\end{tabular}
\caption{Details of the \hii\ spectral cubes. The columns are the maximum UV distance (in k$\lambda$) used for the imaging, the angular resolution, the position angle, the RMS noise per 12.5~km~s$^{-1}$ channel, and the 3$\sigma$ H{\sc i}  column density sensitivity for a line width of 12.5~km~s$^{-1}$.}
\label{table:cubes}
\end{table}

The GMRT Band-5 receivers were used to observe SDSS~J0158+6542 in January, February, July, and August 2020, in proposals DDTC115 and 38\_081 (PI: Nissim Kanekar), for a total time of $\approx 45$~hours (total on-source time $\approx 32$~hours). The GMRT Software Backend was used as the correlator, with a bandwidth of 4.167~MHz, sub-divided into 512 channels, and centred at 1374.26~MHz; the total velocity coverage is $\approx 900$~\kms\ at $z = 0.03399$, with a velocity resolution of $\approx 1.8$~km~s$^{-1}$. Observations of the standard calibrators 3C48 and 3C147 at the start and the end of each run were used to calibrate the flux density scale, and of the nearby compact source J0157+7442, to calibrate the complex antenna gains and bandpass shapes.

The data were analysed in the Common Astronomy Software Applications package \citep[{\sc casa} version 5.6;][]{McMullin07}, following standard procedures. For each observing run, we made a single-channel visibility data set and performed initial editing to elide visibilities from non-working antennas and visibilities affected by radio-frequency interference (RFI). The antenna-based complex gains were estimated from this single-channel data set, using the routine {\sc gaincalR} \citep{Chowdhury20}. The edits and gains were then applied to all channels of the multi-channel data set, and the antenna-based bandpass shapes were then estimated using the routine {\sc bandpassR} \citep{Chowdhury20}. The gains and bandpasses were then applied to the visibilities of SDSS~J0158+6542, and the calibrated visibilities from all epochs combined to produce a single data set. The calibrated visibilities were then spectrally averaged, excluding edge channels and line channels, to produce a ``channel-0'' continuum data set, with a spectral resolution of $\approx 0.5$~MHz (to avoid bandwidth smearing). An iterative imaging and self-calibration procedure was then run on the channel-0 visibilities, to more accurately determine the antenna-based gains. The self-calibration solved for the phases of the complex gains alone, as the signal-to-noise ratio of the data was insufficient to accurately solve for both the relative amplitudes and the phases; the gain amplitudes  were hence fixed to the values obtained from the original calibration. The imaging and self-calibration procedure, along with editing of the residual visibilities, was carried out until no improvement was seen in either the image or the residual visibilities after further iterations. The final continuum image has a synthesized beam of $2.1'' \times 1.9''$ and a root mean square (RMS) noise of $\approx 50~\mu$Jy/beam, away from bright continuum sources. After the self-calibration, the refined antenna-based gains were applied to the spectral-line visibility data set, and the routine {\sc uvsub} was used to subtract out all continuum emission from the calibrated visibilities. The residual visibilities were then inspected visually, and any data affected by low-level RFI were removed.

The hybrid array configuration of the GMRT \citep{Swarup1991} allowed us to make \hii\ images with a wide range of angular resolutions.  Our search for \hii\ emission from SDSS~J0158+6542 was carried out by imaging the residual calibrated visibilities at a velocity resolution of 12.5~\kms, to produce spectral cubes at angular resolutions of $\approx 3'' - 50''$. This provided measurements of the total H{\sc i} content of SDSS~J0158+6542 and its H{\sc i} spatial and kinematic distributions, using coarse and fine angular resolutions, respectively. Before making the spectral cube at any given resolution, we first made a continuum image from the residual visibilities of the off-line channels (also excluding the edge channels), and subtracted out any detected continuum emission from the full visibility data set. This was done in order to remove any extended continuum emission that was not detected in our original $2.1'' \times 1.9''$-resolution continuum image. The cubes were made using the routine {\sc tclean} with the Briggs robust parameter \citep{Briggs95} set to $1$, in the barycentric frame. The cubes were ``cleaned'' during the imaging process, down to a threshold of $\approx 0.5\sigma$, where $\sigma$ is the RMS noise on each 12.5~km~s$^{-1}$ plane of the cube. Finally, for each cube, the routine {\sc imcontsub} was used to fit, and subtract out, a first-order polynomial to line-free regions (again excluding edge channels), in order to remove any residual spectral baseline. Table~\ref{table:cubes} summarizes the details of the different spectral cubes.

Next, the spatial and kinematic distributions of \hi\ in galaxies can be studied by taking velocity moments of the 3-dimensional spectral cubes, integrating over the velocity axis to obtain 2-dimensional images. The zeroth moment image yields the velocity-integrated H{\sc i} 21~cm emission at each spatial location, while the first moment image yields the H{\sc i}  velocity field, weighted by the \hii\ line intensity. The moment images were made with the routine {\sc momnt} in the Astronomical Image Processing System \citep[{\sc aips}; ][]{Greisen2003} package, producing \hi\ images and velocity fields of SDSS~J0158+6542 and its environs at angular resolutions of $\approx 3'' - 50''$. While making the moment images, each cube was first visually inspected to identify the velocity range of \hii\ emission, between nulls; this was used to select the velocity range over which the cubes were integrated. Next, in order to not be biased by noise peaks in the spectral cubes, we used a flux density threshold for the inclusion of putative \hii\ emission signals in the moment images. The threshold was applied after smoothing the cube in velocity (Hanning smoothing to a resolution of $25$~km~s$^{-1}$) and position [smoothing by a Gaussian with full width at half maximum $\approx (1.0-1.7) \times$ the synthesized beam]. A threshold of $\approx 1.5\sigma - 2\sigma$ was chosen, after applying the above smoothing, where $\sigma$ is the RMS noise per 12.5~km~s$^{-1}$ channel for each spectral cube. We emphasize that the smoothing was only carried out in order to mask the original cubes, and that the results do not depend significantly on the choice of the threshold. For the velocity fields, noise spikes were further excluded by blanking the final images at locations that did not show a clear detection of \hii\ emission in the \hi\ intensity image.

\section{Results and Discussion}
\label{sec:results}

\begin{figure*}
\centering
\includegraphics[width=\linewidth]{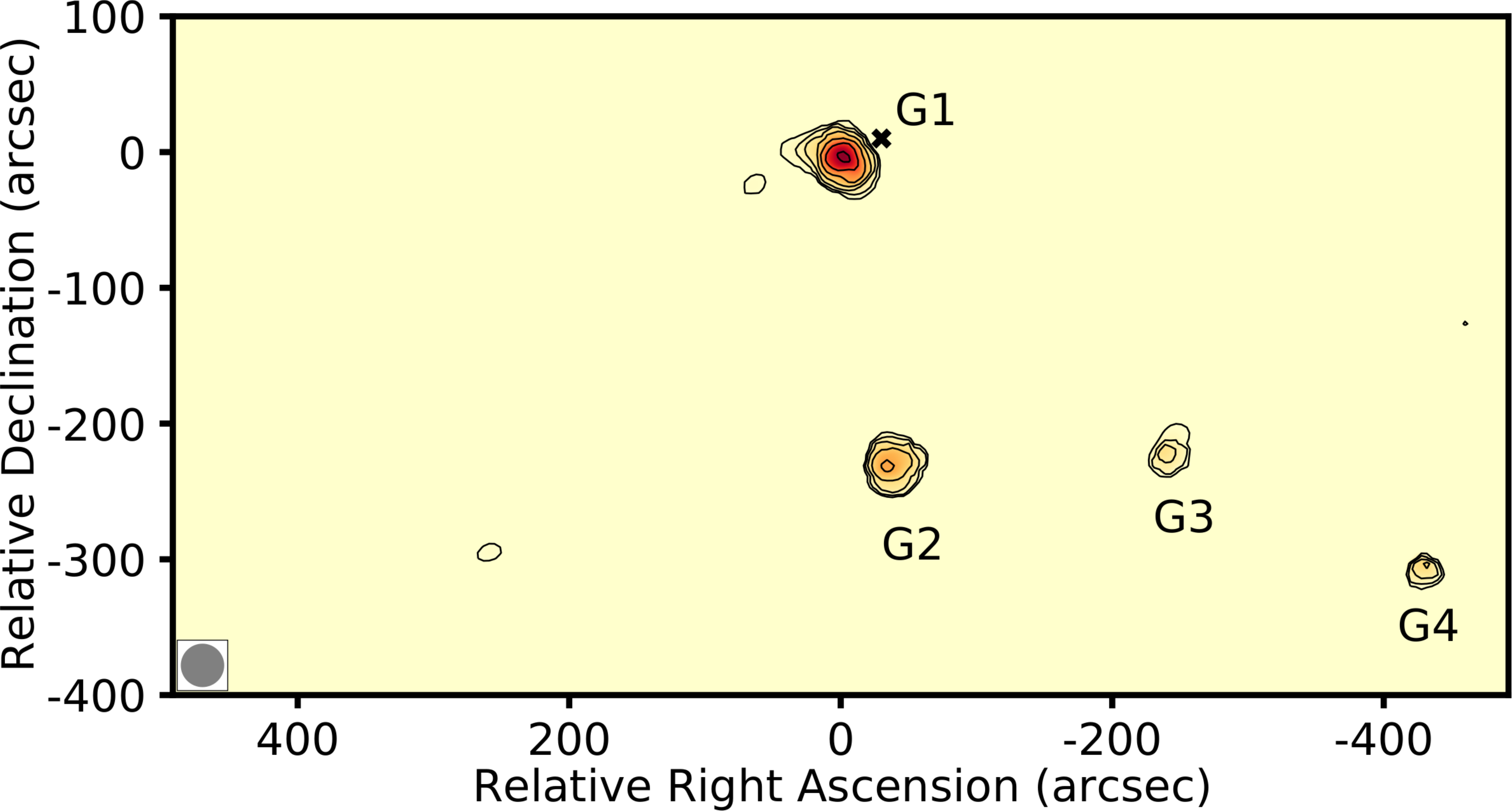}
	\caption{A wide-field \hii\ emission image of the field of SDSS~J0158+6542. The axes co-ordinates are in arcseconds, relative to the FRB position 
	\citep[01h58m00.75s, +65d43'00.315''; ][]{Marcote2020}. The image has an angular resolution of $32.0'' \times 32.0''$, and was obtained by integrating the velocity range $-105$~\kmps\ to $+169$~\kmps, where \hii\ emission is detected in the cube. \hii\ emission is clearly detected from SDSS~J0158+6542, and three galaxies to the south and south-west, labelled G2, G3, and G4. The \hii\ emission of G1 is too faint to be clearly detected at the relatively low sensitivity of this image; we have indicated the position of G1, identified from the higher-resolution images of Figs.~\ref{fig:9arc}[A] and \ref{fig:highres}[A], with a black cross.}
\label{fig:lowres}
\end{figure*}

The large-scale H{\sc i} spatial distribution of the field of SDSS~J0158+6542, at an angular resolution of $32.0'' \times 32.0''$, is shown in Fig.~\ref{fig:lowres}. This was obtained by integrating the \hii\ emission over the velocity range $ -105$~km~s$^{-1}$ to $+169$~km~s$^{-1}$, to cover the velocity range of the \hii\ emission detected from all galaxies in the field.  Besides the \hii\ emission of SDSS~J0158+6542, emission is also detected from three galaxies to the south and south-west, indicated as G2, G3, and G4 in the figure, all $\gtrsim 3.8'$, i.e. $\gtrsim 160$~kpc, away from SDSS~J0158+6542. The \hii\ emission of SDSS~J0158+6542 is seen to be resolved in Fig.~\ref{fig:lowres}, even at the coarse angular resolution of $32'' \times 32''$. The emission extends over a region of angular size $\approx 35''$; a similar angular extent is obtained from the image made at a resolution of $15.5'' \times 15.5''$ (not shown here).

\begin{figure*}
\centering
\includegraphics[width=\linewidth]{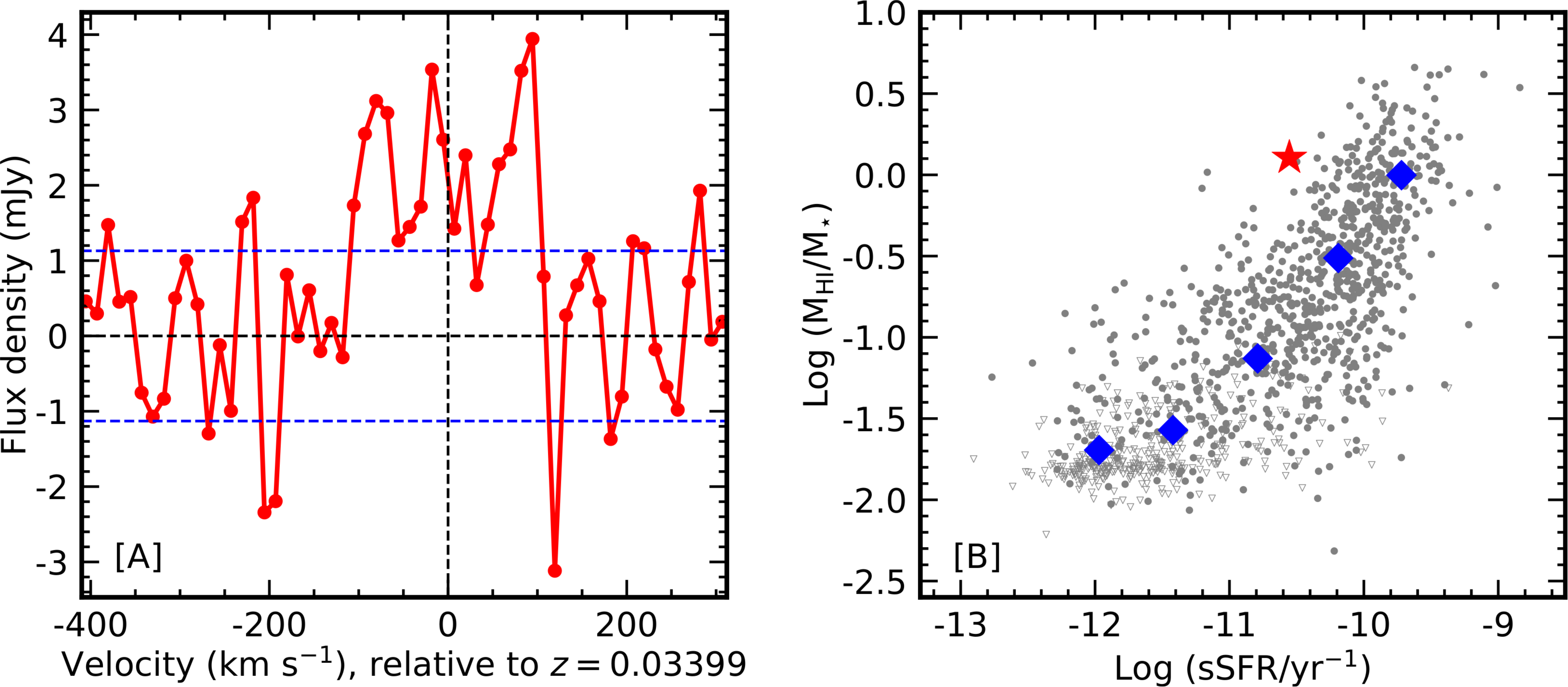}
\caption{[A]~The \hii\ spectrum of SDSS~J0158+6542 at a velocity resolution of 12.5~km~s$^{-1}$, obtained from the spectral cube of angular resolution $32.0'' \times 32.0''$; the blue dashed lines indicate the $\pm 1\sigma$ errors on the spectrum. The x-axis is velocity, in \kms, relative to the galaxy redshift of $z = 0.03399$ \citep{Marcote2020,Tendulkar2021}. [B]~ The inferred \hi-to-stellar mass ratio, plotted against the specific star formation rate. The red star shows the measurements for SDSS~J0158+6542. The grey symbols (\hii\ emission detections in circles and non-detections in open triangles) show data for the xGASS galaxies \citep{Catinella2018}, with the blue diamonds indicating the weighted medians of the logarithm of the H{\sc i}-to-stellar mass ratio for five bins of log$\rm \left[sSFR/yr^{-1}\right]$ in the xGASS sample \citep{Catinella2018}.  The H{\sc i}-to-stellar mass ratio of SDSS~J0158+6542 is approximately an order of magnitude higher than the median xGASS \hi-to-stellar mass ratio at the same sSFR, and the ratio for SDSS~J0158+6542 lies above the envelope of xGASS values for comparable values of the sSFR.
\label{fig:himass}}
\end{figure*}

\begin{figure*}
\centering
\includegraphics[width=\linewidth]{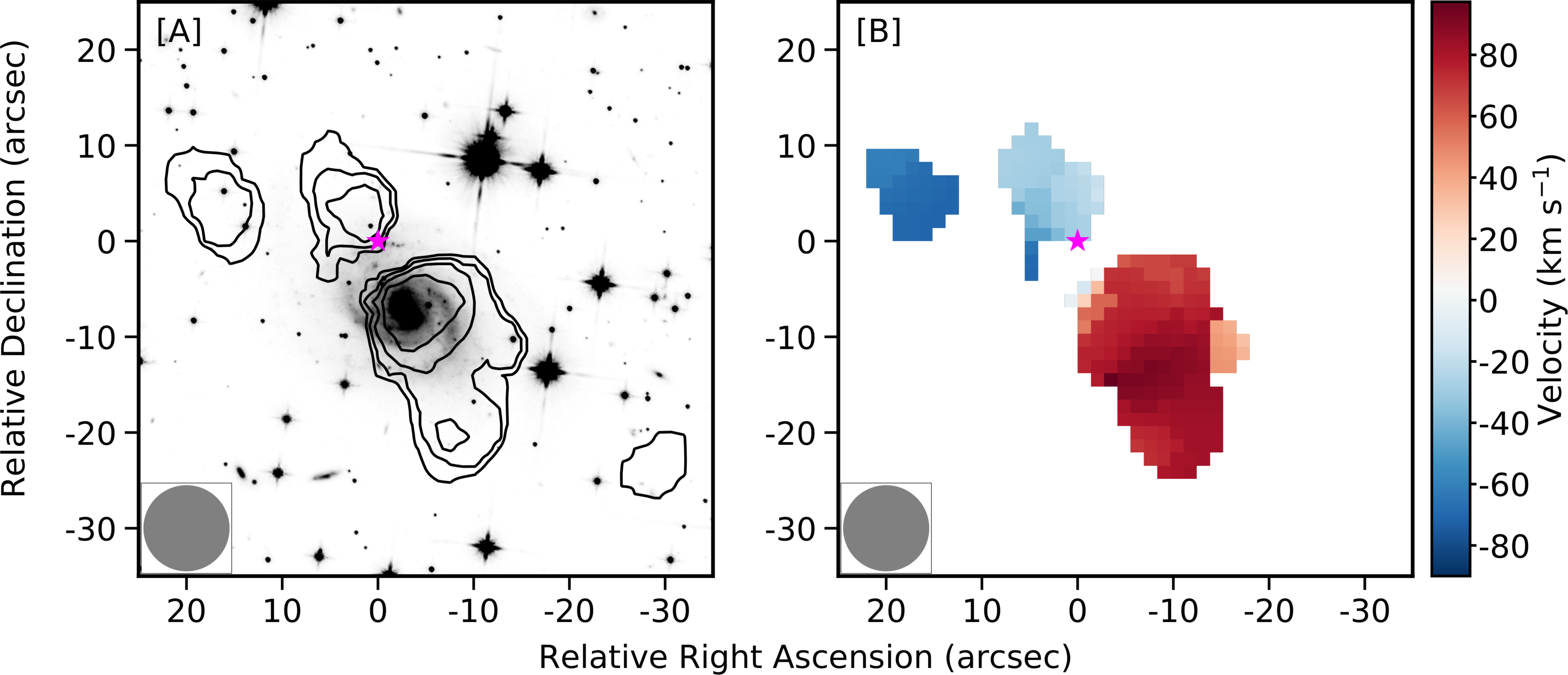}
	\caption{The [A]~\hi\ spatial distribution and [B]~\hi\ velocity field of SDSS~J0158+6542, at an angular resolution of $9.0''\times 9.0''$. The axes co-ordinates are in arcseconds, relative to the FRB position 
	\citep[01h58m00.75s, +65d43'00.315''; ][]{Marcote2020}, indicated by the magenta star in each panel. The H{\sc i} spatial distribution is overlaid on the HST F110W image \citep{Mannings2021,Tendulkar2021} of the field of SDSS~J0158+6542. The \hii\ intensity image was obtained by integrating over the velocity range $-105$~km~s$^{-1}$ to $+107$~km~s$^{-1}$ which shows \hii\ emission in the low-resolution spectrum of Fig.~\ref{fig:himass}[A].  The first contour is at an \hi\ column density of $2.7 \times 10^{20}$~cm$^{-2}$, with successive contours increasing by a factor of $\sqrt{2}$ in \hi\ column density. The lack of H{\sc i} between the centre of SDSS~J0158+6542 and the FRB location is clearly visible. The velocity field in [B] shows that the extended emission to the north and north-east of SDSS~J0158+6542 is blueshifted relative to $z=0.03399$, while the emission that extends south of the galaxy lies redward of the systemic velocity.
\label{fig:9arc}}
\end{figure*}

\begin{figure*}
\centering
\includegraphics[scale=0.145]{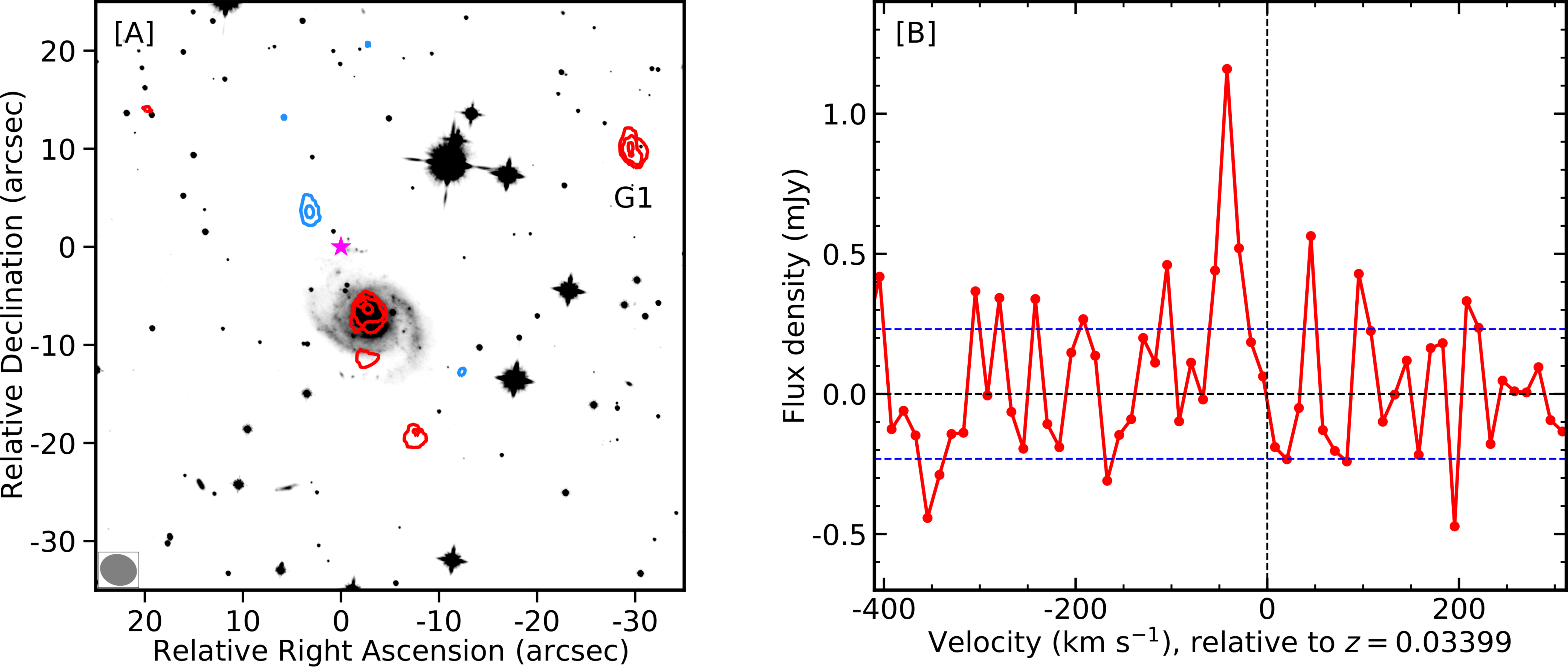}
	\caption{[A]~The  H{\sc i} spatial distribution in SDSS~J0158+6542 and surroundings, at an angular resolution of $3.8'' \times 3.2''$.
	The axes co-ordinates are in arcseconds, relative to the FRB position 
	\citep[01h58m00.75s, +65d43'00.315''; ][]{Marcote2020}, indicated by the magenta star. The \hi\ intensity image is a composite of two images, made by separately integrating over the two velocity ranges, $-55.5 \ {\rm to} \ +32.0$~km~s$^{-1}$ (blue contours), and $+47.5 \ {\rm to} \ +132.0$~km~s$^{-1}$ (red contours). The first contour is at an \hi\ column density of $1.6 \times 10^{21}$~cm$^{-2}$, with successive contours increasing by a factor of $\sqrt{2}$ in \hi\ column density. \hii\ emission is clearly detected at the galaxy centre, from the \hi\ clump next to the FRB location, and from the galaxy G1. [B]~The \hii\ emission spectrum from the \hi\ clump next to the FRB location at a resolution of $3.8'' \times 3.2''$; this has an \hi\ column density of $(2.41 \pm 0.45) \times 10^{21}$~cm$^{-2}$. The blue dashed lines indicate the $\pm 1\sigma$ errors on the spectrum. The x-axis is velocity, in \kms, relative to the galaxy redshift of $z = 0.03399$ \citep{Marcote2020,Tendulkar2021}.
\label{fig:highres}}
\end{figure*}

Fig.~\ref{fig:himass}[A] shows the \hii\ spectrum of SDSS~J0158+6542, obtained from the spectral cube at a resolution of $32'' \times 32''$. The \hii\ emission is detected at $\approx 8.3\sigma$ significance and is extended over a velocity range of $\approx 225$~km~s$^{-1}$ (between nulls), with a velocity-integrated line flux density of $(0.501 \pm 0.060)$~Jy~km~s$^{-1}$. This implies an \hi\ mass of $\rm M_{H{\textsc i}} = (2.74 \pm 0.33) \times 10^9 \, M_\odot$. The star-formation rate (SFR) and stellar mass of SDSS~J0158+6542 are $(0.06 \pm 0.02) \, \Msun$~yr$^{-1}$ and $\rm M_\star = (2.15 \pm 0.33) \times 10^9 \, M_\odot$, respectively \citep{Heintz2020}; this gives a specific star-formation rate (sSFR) of $(0.028 \pm 0.010) \times 10^{-9}$~per~year.  Combining the stellar mass and the H{\sc i} mass, we obtain  an H{\sc i}-to-stellar mass ratio of $\rm f_{H\textsc{i}} \equiv \rm M_{H\textsc{i}}/M_\star \approx 1.3$.

Fig.~\ref{fig:himass}[B] compares the H{\sc i}-to-stellar mass ratio of SDSS~J0158+6542 with that of galaxies (at similar redshifts) of the stellar-mass-selected xGASS sample \citep{Catinella2018}. The figure shows that xGASS galaxies with sSFR values comparable to that of SDSS~J0158+6542 have median H{\sc i}-to-stellar mass ratios $\rm f_{H\textsc{i}} \approx 0.1$, lower than that of SDSS~J0158+6542 by an order of magnitude. Indeed, the ratio for SDSS~J0158+6542 lies above the envelope of the ratios for the xGASS galaxies with comparable sSFR values, indicating that SDSS~J0158+6542 is a gas-rich galaxy. Further, SDSS~J0158+6542 lies well below the star-forming main sequence, and is almost a quiescent galaxy \citep{Bhandari2021}. The combination of gas-richness and near-quiescence indicates that SDSS~J0158+6542 has acquired significant amounts of \hi\ in the recent past.

Next, to characterize the spatial and velocity distribution of the H{\sc i} in SDSS~J0158+6542, we mapped the \hii\ emission at an angular resolution of $9'' \times 9''$; this provides $\approx 4$ independent beams across the galaxy, with a spatial resolution of $\approx 6.3 \ {\rm kpc} \times 6.3 \ {\rm kpc}$, while retaining a high \hi\ column density sensitivity.  Fig.~\ref{fig:9arc}[A] shows the velocity-integrated \hii\ image of SDSS~J0158+6542 at this resolution, overlaid on the Hubble Space Telescope (HST) F110W image \citep{Tendulkar2021,Mannings2021}, while Fig.~\ref{fig:9arc}[B] shows the \hii-intensity weighted \hi\ velocity field; the FRB location is indicated by the magenta star in each panel.

Fig.~\ref{fig:9arc}[A] shows that the \hii\ emission arises from well beyond the optical extent of the galaxy, towards the north-east and the south. The \hii\ emission breaks up into three distinct components, one extending southwards from the optical galaxy centre, and the other two to the north and north-east of the optical galaxy, with the northern emission abutting on the FRB position. The \hi\ spatial distribution, with an \hi\ tail to the north-east and a counter-tail to the south, is reminiscent of minor merger galaxies of the M51 type \citep{Rots1990,Verheijen2001,Hibbard2001}. A clear hole is seen in the H{\sc i} distribution between the galaxy centre and the FRB location; the $3\sigma$ upper limit on the H{\sc i} column density in this region is $1.6 \times 10^{20}$~cm$^{-2}$ per 12.5~km~s$^{-1}$ channel. The contours of the \hii\ emission around the galaxy centre lie close together towards the north, but are extended towards the south, indicating that the \hi\ in the northern part of the galaxy has been compressed. 

The intensity-weighted \hi\ velocity field of SDSS~J0158+6542 of Fig.~\ref{fig:9arc}[B] shows that the gas velocities are blueshifted to the east and north-east of the FRB location, but redshifted south of the FRB location \citep[relative to the galaxy redshift $z = 0.03399$; ][]{Tendulkar2021}. The \hi\ velocity field is very different from that expected from a rotating disk galaxy, and, in fact, shows no evidence of regular rotation.  The velocity increases along the \hi\ tail to the north-east of SDSS~J0158+6542, as is typical in minor merger tails \citep[e.g.][]{Rots1990}.

Moving to even finer resolution, the $3.8'' \times 3.2''$ resolution image of Fig.~\ref{fig:highres}[A] is a composite, made by separately integrating over the two velocity ranges, $-55.5 \ {\rm to} \ +32.0$~km~s$^{-1}$ (blue contours), and $+47.5 \ {\rm to} \ +132.0$~km~s$^{-1}$ (red contours). This was done because the \hii\ emission in the detected regions shows a clear separation in velocity, and the signal-to-noise ratio in each region can hence be increased by narrowing the velocity range to only pick out its \hii\ emission. Fig.~\ref{fig:highres}[A] shows that, at a resolution of $\approx 3.8'' \times 3.2''$, the \hi\ arises from compact components at the centre of the optical galaxy and close to the FRB location. Further, a new compact component, from the companion G1, is now visible $\approx 31.7''$ to the north-west of SDSS~J0158+6542. Fig.~\ref{fig:highres}[B] shows the \hii\ emission spectrum from the \hi\ clump close to (within $3.5$~kpc of) the FRB location; this has a very high H{\sc i} column density of $(2.41 \pm 0.45) \times 10^{21}$~cm$^{-2}$, within a factor of $\approx 2$ of the \hi\ column density at the centre of the galaxy. This H{\sc i} clump lies adjacent to, and to the north-east of, the V-shaped knot of star-formation detected in the HST image \citep{Marcote2020,Tendulkar2021}. It has been suggested \citep{Marcote2020} that this star-forming knot might have arisen due to an interaction, either within SDSS~J0158+6542 or between the galaxy and a dwarf companion. 

Taken together, our results suggest that SDSS~J0158+6542 has recently undergone a minor merger with a gas cloud or a galaxy, that resulted in an increase in its gas content, disturbed the H{\sc i} in the galaxy disk, compressed the H{\sc i} near the FRB location, and yielded a distorted velocity field. It is likely that the increased H{\sc i} column density triggered a burst of star-formation that gave rise to both the FRB progenitor and the star-forming knot seen in the optical image \citep{Marcote2020,Tendulkar2021,Mannings2021}. 

The FRB site lies between the star-forming knot detected in H$\alpha$ emission \citep{Tendulkar2021} and the location of high \hi\ column density in Fig.~\ref{fig:highres}[A]. We tentatively detect (at $\approx 3.6\sigma$ significance) \hii\ emission from the FRB location, obtaining an \hi\ column density of $(1.32 \pm 0.37) \times 10^{21}$~\cm. However, the angular separation between the FRB site and the star-forming knot is only $\approx 0.35''$, an order of magnitude smaller than our angular resolution of $3.8'' \times 3.2''$; it is hence possible that the detected \hii\ emission arises from the star-forming knot, rather than from the FRB location. A low \hi\ column density at the FRB location would be consistent with the low free-free optical depth inferred from the detections of bursts at frequencies of $100-300$~MHz \citep{Chawla2020,Pleunis2021}, and with the non-detection of H$\alpha$ emission at the FRB location \citep{Tendulkar2021}. These suggest that the FRB site has a low gas column density and is hence unlikely to be a star-forming region; the FRB source appears likely to have moved to its present location from the star-forming knot detected in the HST image \citep{Tendulkar2021}.

\begin{figure}
\centering
\includegraphics[width=0.9\linewidth]{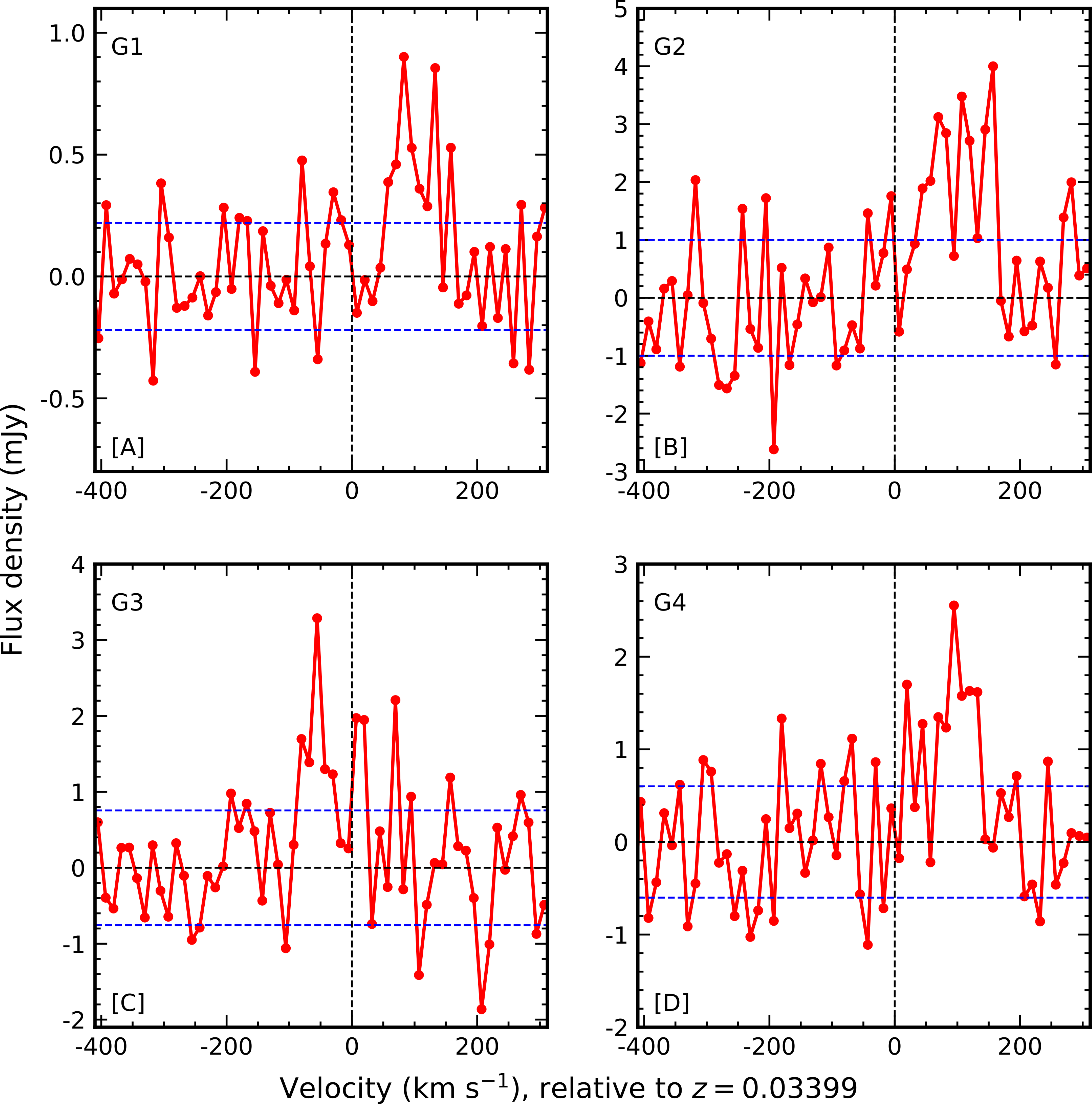}
\caption{\hii\ spectra of the companions G1, G2, G3, and G4. The \hii\ spectra have a velocity resolution of 12.5~km~s$^{-1}$. The spectrum of G1 was obtained from the spectral cube with an angular resolution of $3.8'' \times 3.2''$, while the remaining three spectra were obtained from the cube with a resolution of $32.0'' \times 32.0''$. The blue dashed lines in each panel indicate the $\pm 1\sigma$ errors on the spectra. 
\label{fig:companions}}
\end{figure}

\begin{table}
\centering
\begin{tabular}{|c|c|c|c|c|}
\hline
\hline
Galaxy	& J2000 co-ordinates &  $\int {\rm S} \ {\rm dV}$ &  $\Delta$V &  M$_{\textrm{H}\textsc{i}}$ \\ 
	&                    &  Jy~km~s$^{-1}$            & km~s$^{-1}$ &    $ \Msun$ \\
\hline
J0158+6542      & 01h58m00.28s, +65d42$'$53.10$''$ & $ 0.501 \pm 0.060$ & $225$ & $(2.74 \pm 0.33) \times 10^9$ \\
\hline
	G1      & 01h57m55.94s, +65d43$'$10.16$''$ & $ 0.0533 \pm 0.0082$ & $112$ & $(2.92 \pm 0.45) \times 10^8$ \\
\hline
	G2      & 01h57m55.53s, +65d39$'$08.53$''$ & $ 0.321 \pm 0.042$ & $137$ & $(1.76 \pm 0.23) \times 10^9 $ \\
\hline
	G3      & 01h57m22.15s, +65d39$'$12.28$''$ & $ 0.171 \pm 0.030$ & $125$ & $(9.4 \pm 1.6) \times 10^8$ \\
\hline
	G4      & 01h56m52.21s, +65d37$'$50.87$''$ & $ 0.164 \pm 0.024$ & $125$ & $(9.0 \pm 1.3) \times 10^8$ \\
\hline
\hline
\end{tabular}
\caption{Details of the \hii\ emission from SDSS~J0158+6542 and the four companions. The columns are the galaxy name, its J2000 co-ordinates, the velocity-integrated \hii\ line flux density, the width of the \hii\ emission between nulls, and the inferred \hi\ mass.
 The spectrum of G1 was obtained from the spectral cube with an angular resolution of $3.8'' \times 3.2''$, while the remaining four spectra were obtained from the cube with a resolution of $32.0'' \times 32.0''$.
\label{table:companions}}
\end{table}

We measured the integrated the \hii\ line flux densities of the companions G2, G3, and G4 from the spectral cube with an angular resolution of $32.0'' \times 32.0''$, and of G1 from the higher-sensitivity cube with an angular resolution of $3.8'' \times 3.2''$; the \hii\ spectra are shown in Fig.~\ref{fig:companions}. In all cases, we also measured the integrated \hii\ line flux densities at coarser resolutions, to ensure that we do not resolve out any spatially-extended emission. The positions of the galaxies (including SDSS~J0158+6542), and their integrated \hii\ line flux densities, \hii\ line velocity widths, and \hi\ masses are listed in Table~\ref{table:companions}.

Our present data do not definitively identify the counterpart of SDSS~J0158+6542 in the minor merger. It is tempting to identify G1, by far the closest of the \hi\ companions of SDSS~J0158+6542 (only $\approx 22$~kpc away), as the merger companion. However, the mass of G1 is relatively low, $\approx 2.9 \times 10^8 \ \Msun$, a factor of $\approx 10$ lower than that of SDSS~J0158+6542. A merger with G1 is thus unlikely to have significantly increased the \hi-to-stellar mass ratio of the FRB host galaxy. Further, no optical emission is detected from the location of G1 in the HST F110W image \citep{Mannings2021,Tendulkar2021}; this suggests that G1 may be an \hi\ cloud in the circumgalactic medium of SDSS~J0158+6542, rather than a companion galaxy. However, the \hi\ mass of G1 is significantly larger than that of the high-velocity cloud population in the Milky Way \citep{Putman12,Westmeier18}, making it unlikely that G1 is a normal high-velocity cloud. It is possible that G1 consists of gas pushed out of SDSS~J0158+6542 by the minor merger; this would explain both the lack of a stellar counterpart and the higher \hi\ mass than that of typical high-velocity clouds.

An estimate of the dynamical mass of the FRB host may be obtained from the \hii\ spectrum of Fig.~\ref{fig:himass}[A], by assuming  that the \hi\ lies in a disk and that the velocity spread arises due to rotation. Using the optical inclination, $i = 33^\circ$ \citep{Tendulkar2021}, we obtain a maximum rotation velocity, $\rm v_{circ} \approx FWHM/(2\times \sin{i}) \approx 185$~\kmps. The angular extent of the \hii\ emission is $\approx 35''$, implying a radius of $\approx 12.5$~kpc. For a rotating disk, we then obtain a dynamical mass of $\approx 10^{11} \ \Msun$ \citep[e.g.][]{Yu2020}. This estimate is likely to be too high by a factor of a few, as the above estimates of both the radius and the maximum rotation velocity contain non-rotational contributions.

Similar disturbed H{\sc i} spatial distributions have been identified in the host galaxies of a nearby gamma ray burst GRB\,980425 \citep{Arabsalmani2019} and a fast blue optical transient AT2018cow \citep{Roychowdhury2019}. In both cases, high H{\sc i} column densities were found close to the location of the transient, akin to the situation for FRB\,20180916B. This suggests that, similar to long-duration gamma-ray bursts and fast optical transients, the progenitor of the source of FRB\,20180916B is likely to be a massive star, formed due to the merger event. The birth of the progenitor of the FRB source in a recent merger event is consistent with results from earlier studies of FRB\,20180916B that suggest that the FRB arises in a neutron star in a high-mass X-ray binary system \citep{Chime2020,Nimmo2021b,Tendulkar2021,Pleunis2021}.

\begin{acknowledgements}
We thank the staff of the GMRT who have made these observations possible. The GMRT is run by the National Centre for Radio Astrophysics of the Tata Institute of Fundamental Research. B.K. and N.K. acknowledge support from the Department of Atomic Energy, under project 12-R\&D-TFR-5.02-0700.  J.X.P., as member of the Fast and Fortunate for FRB  Follow-up team, acknowledges support from  NSF grant AST-1911140. We thank Alexandra Mannings for providing us with the HST F110W image of the field of SDSS~J0158+6542. We also thank Jacqueline van Gorkom for discussions, Aditya Chowdhury and Jayaram Chengalur for comments on an earlier draft, and an anonymous referee for suggestions that improved the manuscript.
\end{acknowledgements}
\bibliographystyle{aasjournal}

\end{document}